\documentclass[twocolumn]{aastex61}

\usepackage{graphicx}
\usepackage{natbib}

\received{July 1, 2016}
\revised{September 27, 2016}
\accepted{\today}
\submitjournal{ApJ}

\shorttitle{Hydrocarbons in OMC--2 FIR\,4}
\shortauthors{Favre et al.}

\begin{document}

\title{SOLIS IV. Hydrocarbons in the OMC--2 FIR\,4 region, a probe of
  energetic particle irradiation of the region\footnote{Based on
    observations carried out under project number L15AA with the IRAM
    NOEMA Interferometer.  IRAM is supported by INSU/CNRS (France),
    MPG (Germany) and IGN (Spain).}}

\correspondingauthor{C. Favre}
\email{cfavre@arcetri.astro.it}

\author{C. Favre}
\affil{Univ. Grenoble Alpes, CNRS, IPAG, F-38000 Grenoble, France}
\affil{INAF-Osservatorio Astrofisico di Arcetri, Largo E. Fermi 5, I-50125, Florence, Italy}
\author{C. Ceccarelli}
\affil{Univ. Grenoble Alpes, CNRS, IPAG, F-38000 Grenoble, France}
\author{A. L\'opez-Sepulcre}
\affil{Univ. Grenoble Alpes, CNRS, IPAG, F-38000 Grenoble, France}
\affil{Institut de Radioastronomie Millim\'etrique, 300 rue de la Piscine, 38406, Saint-Martin d'H\`eres, France}
\author{F. Fontani}
\affil{INAF-Osservatorio Astrofisico di Arcetri, Largo E. Fermi 5, I-50125, Florence, Italy}
\author{R. Neri}
\affil{Institut de Radioastronomie Millim\'etrique, 300 rue de la
  Piscine, 38406, Saint-Martin d'H\`eres, France}
\author{S. Manigand}
\affil{Univ. Grenoble Alpes, CNRS, IPAG, F-38000 Grenoble, France}
\affil{Centre for Star and Planet Formation, Niels Bohr Institute $\&$ Natural History Museum of Denmark, University of Copenhagen, \O ster Voldgade 5-7, 1350 Copenhagen K., Denmark}
\author{M. Kama}
\affil{Institute of Astronomy, University of Cambridge, Madingley Road, Cambridge CB3 0HA, UK}
\author{P. Caselli}
\affil{Max-Planck-Institut f\"ur extraterrestrische Physik, Giessenbachstrasse 1, 85748  Garching,  Germany}
\author{A. Jaber Al-Edhari}
\affil{Univ. Grenoble Alpes, CNRS, IPAG, F-38000 Grenoble, France}
\affil{University of AL-Muthanna, College of Science, Physics Department, AL-Muthanna, Iraq}
\author{C. Kahane}
\affil{Univ. Grenoble Alpes, CNRS, IPAG, F-38000 Grenoble, France}
\author{F. Alves}
\affil{Max-Planck-Institut f\"ur extraterrestrische Physik, Giessenbachstrasse 1, 85748  Garching,  Germany}
\author{N. Balucani}
\affil{Dipartimento di Chimica, Biologia e Biotecnologie, Universit\`a di Perugia, Via Elce di Sotto 8, I-06123 Perugia, Italy}
\author{E. Bianchi}
\affil{INAF-Osservatorio Astrofisico di Arcetri, Largo E. Fermi 5, I-50125, Florence, Italy}
\author{E. Caux}
\affil{Universit\'e de Toulouse, UPS-OMP, IRAP, Toulouse, France}
\affil{CNRS, IRAP, 9 Av. Colonel Roche, BP 44346, F-31028 Toulouse Cedex 4, France}
\author{C. Codella}
\affil{INAF-Osservatorio Astrofisico di Arcetri, Largo E. Fermi 5, I-50125, Florence, Italy}
\author{F. Dulieu}
\affil{LERMA, Universit\'{e} de Cergy-Pontoise, Observatoire de Paris, PSL Research University, CNRS, Sorbonne Universit\'{e}, UPMC Univ. Paris 06, \'{E}cole normale sup\'{e}rieure, France.}
\author{J. E. Pineda}
\affil{Max-Planck-Institut f\"ur extraterrestrische Physik, Giessenbachstrasse 1, 85748  Garching,  Germany}
\author{I.~R. Sims}
\affil{Institut de Physique de Rennes, UMR CNRS 6251, Universit\'e de Rennes 1, 263 Avenue du G\'en\'eral Leclerc, F-35042 Rennes Cedex, France}
\author{P. Theul\'e}
\affil{Aix-Marseille Universit\'e, PIIM UMR-CNRS 7345, 13397 Marseille, France}

\begin{abstract} 
We report new interferometric images of cyclopropenylidene, c--C$_3$H$_2$, towards the young protocluster OMC--2~FIR\,4. The observations were performed at 82 and 85~GHz with the NOrthern Extended Millimeter Array (NOEMA) as part of the project Seeds Of Life In Space (SOLIS). In addition, IRAM-30m data observations were used to investigate the physical structure of OMC--2~FIR\,4.
We find that the c--C$_3$H$_2$ gas emits from the same region where previous SOLIS observations showed bright HC$_5$N emission. From a non-LTE analysis of the IRAM-30m data, the c--C$_3$H$_2$ gas has an average temperature of $\sim$40 K, a H$_2$ density of $\sim$3$\times$10$^{5}$~cm$^{-3}$, and a c--C$_3$H$_2$ abundance relative to H$_2$ of ($7\pm1$)$\times$10$^{-12}$. In addition, the NOEMA observations provide no sign of significant c--C$_3$H$_2$ excitation temperature gradients across the region (about 3-4 beams), with T$_{ex}$ in the range 8$\pm$3 up to 16$\pm$7~K. We thus infer that our observations are inconsistent with a physical interaction of the OMC--2~FIR\,4 envelope with the outflow arising from OMC--2~FIR\,3, as claimed by previous studies.
The comparison of the measured c--C$_3$H$_2$ abundance with the predictions from an astrochemical PDR model indicates that OMC--2~FIR\,4 is irradiated by a FUV field $\sim$1000 times larger than the interstellar one, and by a flux of ionising particles $\sim$4000 times larger than the canonical value of $1\times10^{-17}$~s$^{-1}$ from the Galaxy cosmic rays, which is consistent with our previous HC$_5$N observations. This provides an important and independent confirmation of other studies that one or more sources inside the OMC--2~FIR\,4 region emit energetic ($\geq10$~MeV) particles.
\end{abstract}

\keywords{ISM: abundances  ---  ISM: clouds -- ISM: molecules -- Radio lines: ISM}


\section{Introduction}
\label{intro}

Earth is so far the only known place where life is present. Why life
emerged and what conditions are essential for that are questions which
challenge our knowledge and still represent a mystery. Very
likely, life is the result of a very long and complex process that
started as early as the formation of the Solar System (hereafter, SS). 
Sparse traces of the process have been left in the SS small bodies \citep[e.g.][]{Caselli:2012}, so that to reconstruct it we need
(also) to look at places that are forming Solar-type planetary systems
today. However, finding such systems depends on the partial knowledge
that we have of the history of the SS formation. In practice,
therefore, reconstructing the SS past history has to be an
``iterative'' process.

Among the information provided by the mentioned SS left traces, two
are particularly relevant for the work presented in this
article. First, the Sun was most likely born in a crowed star cluster in
the vicinity of high-mass stars, and not in an isolated cloud \citep[e.g.][]{Adams:2010}. Second, it underwent a period of intense
irradiation from energetic ($\geq 10$ MeV) particles, even though the
cause is not clear yet \citep[e.g.][]{Gounelle:2013}.

When taking these two facts into account, the source OMC--2 FIR\,4,
north of the famous Orion KL region, is so far the best and closest analogue of
the SS progenitor in our hands. Indeed available observations show
that OMC--2 FIR\,4 is a cluster of several young protostars \citep{Shimajiri:2011,Shimajiri:2015,Lopez-Sepulcre:2013}
 and that it is permeated by a flux of energetic particles, cosmic-ray (CR) like,
which ionise the molecular gas at a rate more than 4000 times the
``canonical'' value of $1\times10^{-17}$ s$^{-1}$ in the Galaxy
(Ceccarelli et al. 2014; Fontani et al. 2017). Given the vicinity of
the Trapezium OB star cluster, the region is also subject to a strong
irradiation from FUV photons, about 1000 times larger than the
interstellar field \citep{Lopez-Sepulcre:2013a}.

For these reasons, OMC--2 FIR\,4 is one of the targets of the project Seeds Of Life In Space \citep[SOLIS;][]{Ceccarelli:2017}
whose goal is to understand how molecular complexity grows in
Solar-type star forming systems. Within this project, we carried out
observations with the IRAM NOrthern Extended Millimeter Array (NOEMA)
interferometer at various frequencies. A first study on the
cyanopolyynes (HC$_3$N and HC$_5$N) showed that carbon chains growth
is favoured in OMC--2 FIR\,4, likely thanks to the presence of the large
CR-like ionising particles flux \citep{Fontani:2017}.

In this work, we present new SOLIS observations of the small
hydrocarbon c-C$_3$H$_2$.  The NOEMA SOLIS data are complemented with
broad band IRAM-30m observations at 1, 2, and 3\,mm. The
article is organised as follows. Section \ref{sec:observations}
describes these new observations. We detected and imaged several lines
as described in Section \ref{results}. With this large and diversified
dataset, we could carry out a sophisticated analysis of the excitation
conditions (Section \ref{sec:temperature-c-c_3h_2}) and the chemical
structure (Section \ref{sec:chemical-modeling}) of the region. In
Section \ref{discussion}, we discuss the information provided by
the new observations and the implications on the processes occurring
in the OMC--2 FIR\,4 region.

\section{Observations and data reduction}
\label{sec:observations}
%
We obtained observations of three c--C$_3$H$_2$ lines with the IRAM
interferometer NOEMA within the SOLIS project \citep[][]{Ceccarelli:2017}. They are here
complemented with IRAM 30m observations of several c--C$_3$H$_2$
lines, detected in the spectral survey previously carried out towards
OMC--2 FIR\,4 in the 3, 2 and 1 mm bands \citep{Lopez-Sepulcre:2015}.
We present the two sets of observations separately.

\subsection{SOLIS NOEMA observations}
Three c--C$_3$H$_2$ lines, one para (2$_{0,2}$--1$_{1,1}$) and two
ortho (3$_{1,2}$--3$_{0,3}$ and 2$_{1,2}$--1$_{0,1}$), were imaged
towards OMC--2 FIR\,4 with the IRAM NOEMA interferometer. The first two
lines, both at $\sim 82$ GHz, were observed with 6 antennas on 2015
August 5, 11, 12, 13 and 19 in the D configuration \citep[see also][]{Fontani:2017}. The third line, at 85 GHz, was observed with 8 antennas on 2016 April 29 and 2016 October 26 in the C configuration.
All three lines were observed with the WideX band correlator, which provides 1843 channels over 3.6 GHz bandwidth with a channel width of 1.95 MHz ($\sim$7.2 km~s$^{-1}$ at 82 GHz).
Table~\ref{tab1} reports the spectroscopic data and the main
characteristics of the observations.

\begin{table}
\caption{\label{tab1} Properties of the cyclopropenylidene
  (c--C$_3$H$_2$) lines observed with the NOEMA interferometer:
  spectroscopic line parameters (transition, frequency, upper level energy and
  Einstein coefficient) and major characteristics of the observations
  (synthesized beam and P.A.). }
\begin{tabular}{llrccr}
\hline\hline
Trans.	&	Freq.	&	E$\rm_{up}$ &   A & beam & P.A.\\
	&(MHz)	&	(K)	& (10$^{-5}$ s$^{-1}$) & ($"$) & ($^o$)\\
\hline
  \multicolumn{4}{l}{para c--C$_3$H$_2$} & &\\ 
2$_{0,2}$--1$_{1,1}$ &	82093.544	&	6.4 	&	2.1	& 9.3$\times$5.9 &-206\\
\hline
  \multicolumn{4}{l}{ortho c--C$_3$H$_2$} & & \\ 
3$_{1,2}$--3$_{0,3}$  &	82966.197	&	16.0	&	1.1	& 9.3$\times$5.9 &-206\\
2$_{1,2}$--1$_{0,1}$  & 85338.896	& 6.4	&	2.6 & 4.7$\times$2.2 & 14\\
\hline
    \hline
\end{tabular}\\
{\small
  We used the spectroscopic data parameters from \citet{Bogey:1986},
  \citet{Vrtilek:1987}, \citet{Lovas:1992} and \citet{Spezzano:2012},
  that are available from the Cologne Database for Molecular
  Spectroscopy molecular line catalog
  \citep[CDMS,][]{Muller:2005}. The Einstein coefficients assume an
  ortho-to-para ratio of 3:1. }
\end{table}

The phase-tracking center was $\alpha_{J2000}$ =
05$^{h}$35$^{m}$26\fs97, $\delta_{J2000}$ =
-05$\degr$09$\arcmin$56$\farcs$8 for all data sets, and the systemic
velocity of OMC2--FIR4 was set to $V_{LSR}=11.4$~km~s$^{-1}$. The primary beams are about 61$\arcsec$ and 59$\arcsec$ for data at 82~GHz and 85~GHz, respectively.
The nearby quasars 3C454.3 and 0524$+$034 were respectively used as bandpass calibrator and gain calibrator for the observations at 82~GHz. Regarding the observations performed at 85~GHz, 0524$+$024 and 0539$-$057 were used as gain (phase and amplitude) calibratiors while 3C454.3 was used as bandpass calibrator. The absolute flux calibration was performed through observations of the quasars LKHA101
(0.21Jy) for 2015 August 5 and 19, MWC349 (1.03 Jy) for 2015
August 11, 12 and 13 and again MWC349 (1.05 Jy) for 2016 observations. 

Continuum subtraction and data imaging were performed using the GILDAS
software\footnote{http://www.iram.fr/IRAMFR/GILDAS/}.  The cleaning of
the spectral lines was performed by using the Hogbom method
\citep{Hogbom:1974}. The resulting synthesized beam size of the molecular emission maps are
9.5\arcsec$\times$6.1\arcsec (P.A.= $-$206$\degr$) and
4.7\arcsec$\times$2.2\arcsec (P.A.= 14$\degr$) at 82 GHz and 85 GHz,
respectively. The NOEMA emission maps shown in this paper are corrected for primary beam.

\subsection{IRAM-30m observations}

Additional observations of the c--C$_3$H$_2$ lines were obtained in
the context of the unbiased spectral survey of OMC--2 FIR\,4 obtained
with the IRAM 30m telescope. The 3 mm (80.5-116.0 GHz), 2 mm
(129.2-158.8 GHz) and 1 mm (202.5-266.0 GHz) bands were observed
between 31 Aug. 2011 and 7 Feb. 2014. The Eight MIxer Receiver (EMIR)
connected to the 195 kHz resolution about 0.7 km s$^{-1}$ at 83~GHz) Fourier Transform Spectrometer
(FTS) units were used. The main beam sizes are about 9-12$\arcsec$, 16$\arcsec$ and 30$\arcsec$ at 1, 2 and 3~mm, respectively.
The observations were carried out in wobbler
switch mode, with a throw of 180$''$. Pointing and focus were
performed regularly.  The coordinates of the IRAM-30m observations are
$\alpha_{J2000}$ = 05$^{h}$35$^{m}$26\fs97 and $\delta_{J2000}$ =
-05$\degr$09$\arcmin$54$\farcs$5. For further details, see \citet{Lopez-Sepulcre:2015}.

We used the package CLASS90 of the GILDAS software
collection to reduce the
data. The uncertainties of calibration are estimated to be lower than
10\% at 3mm and 20\% at 2 and 1mm. After subtraction of the continuum
emission via first-order polynomial fitting, a final spectrum was
obtained by stitching the spectra from each scan and frequency
setting. The intensity was converted from antenna temperature
($T_\mathrm{ant}^\ast$) to main beam temperature ($T_\mathrm{mb}$)
using the beam efficiencies provided at the IRAM web site for the
epoch of the observations.

%
\section{Results}
\label{results}

\subsection{c--C$_3$H$_2$ emission maps}\label{sec:c-c_3h_2-emission}
\begin{figure*}
\centering
\includegraphics[angle=0,width=8cm,]{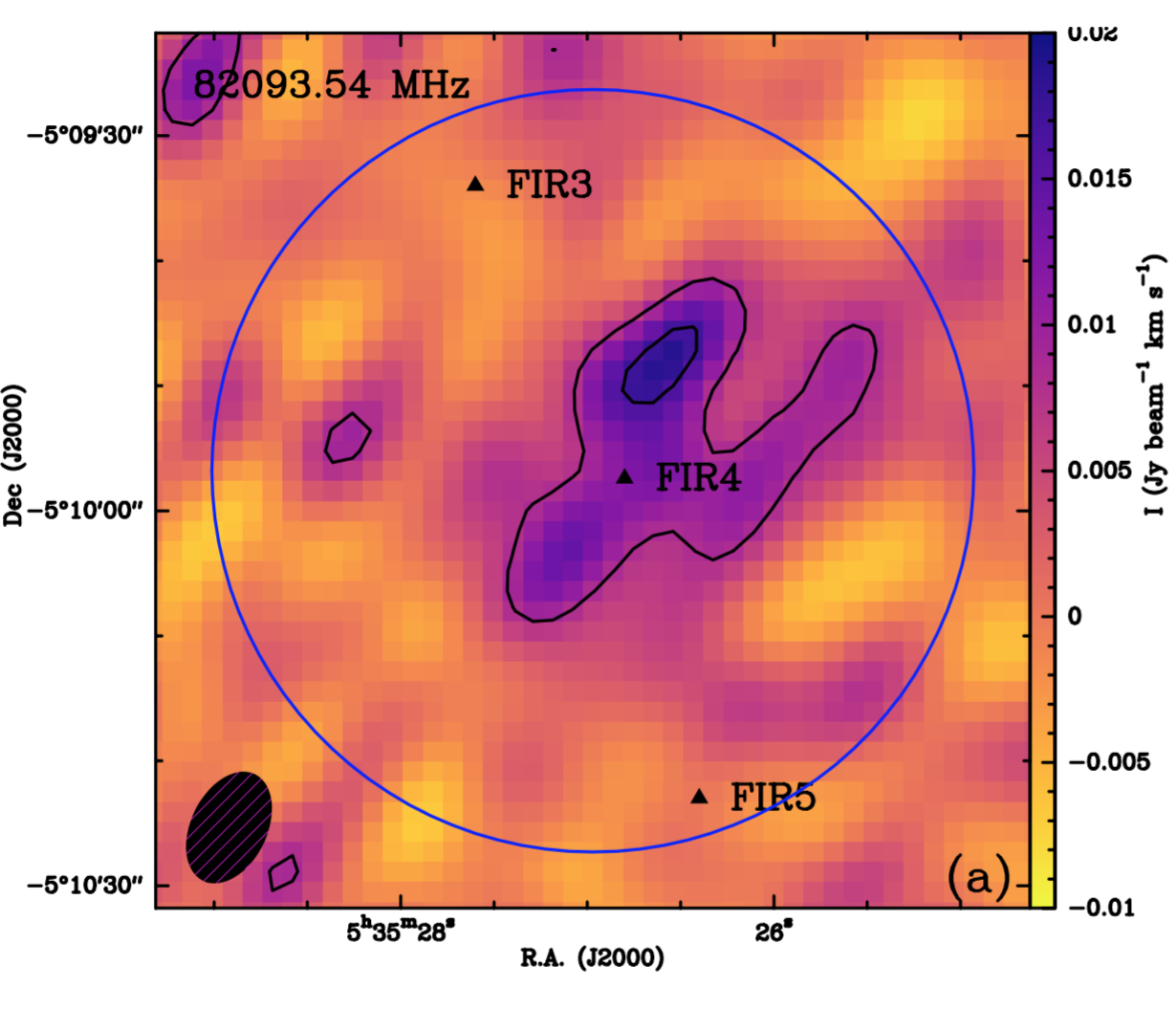}
\includegraphics[angle=0,width=8cm]{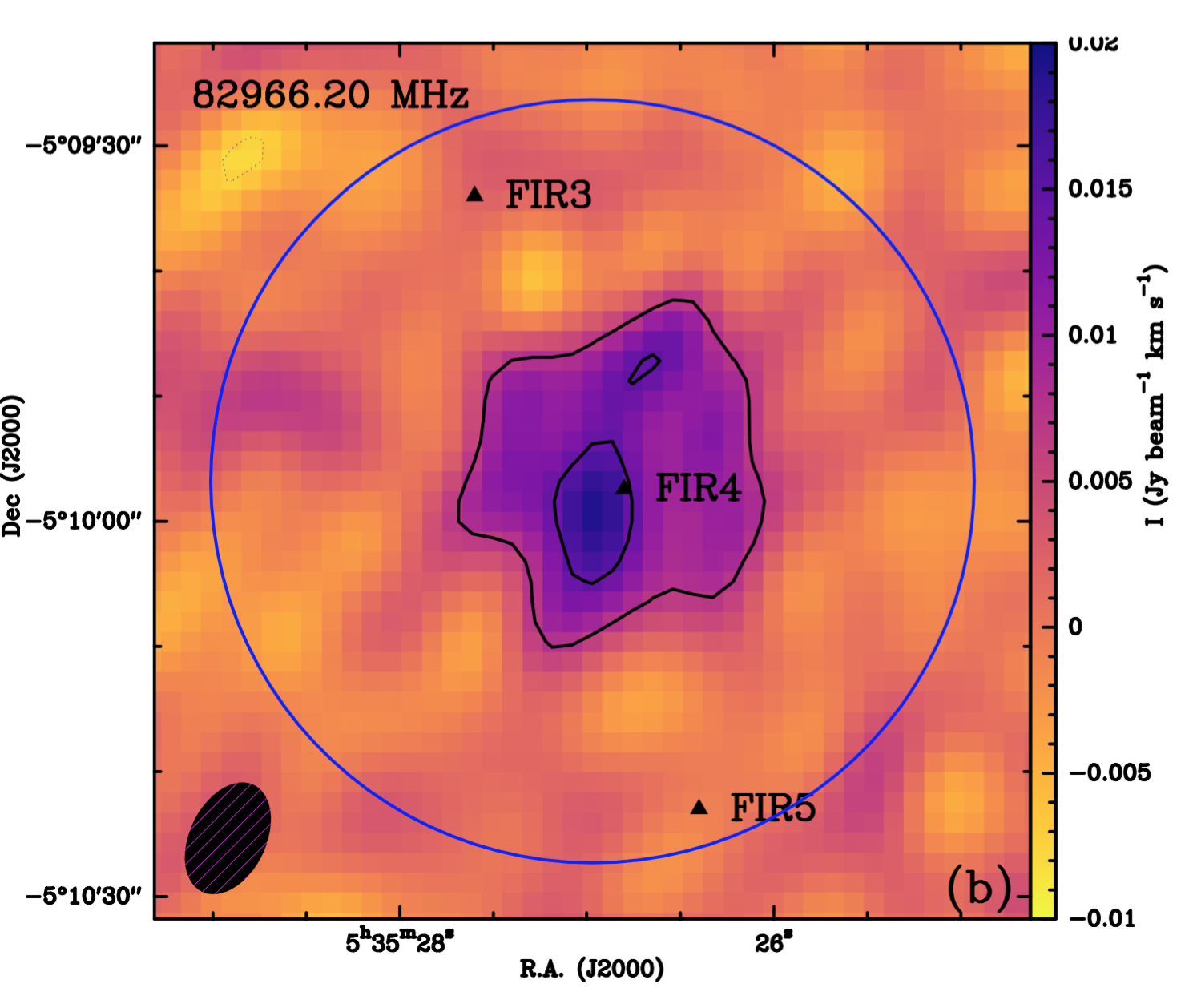}
\includegraphics[angle=0,width=8cm]{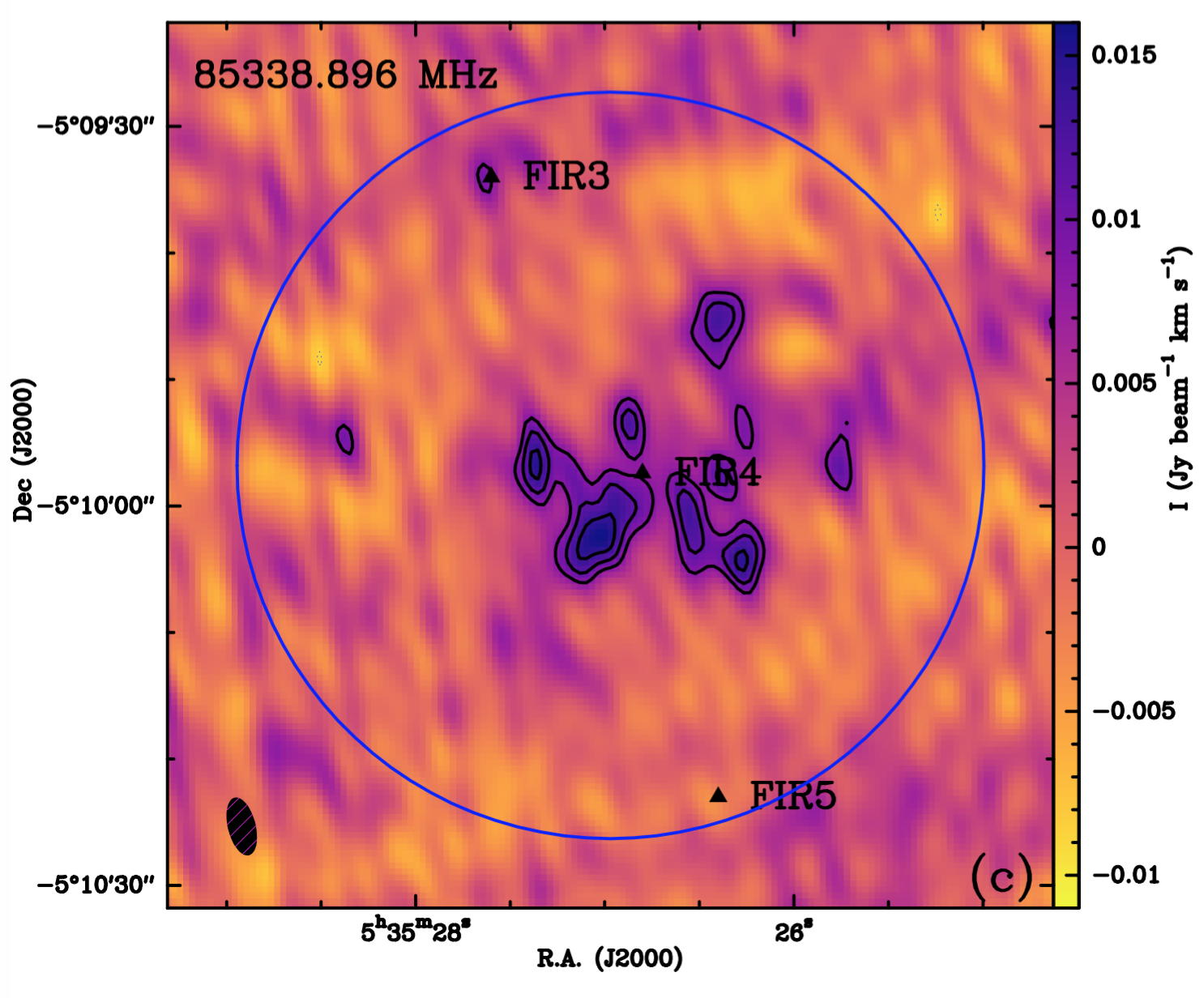}
\includegraphics[angle=0,width=8cm]{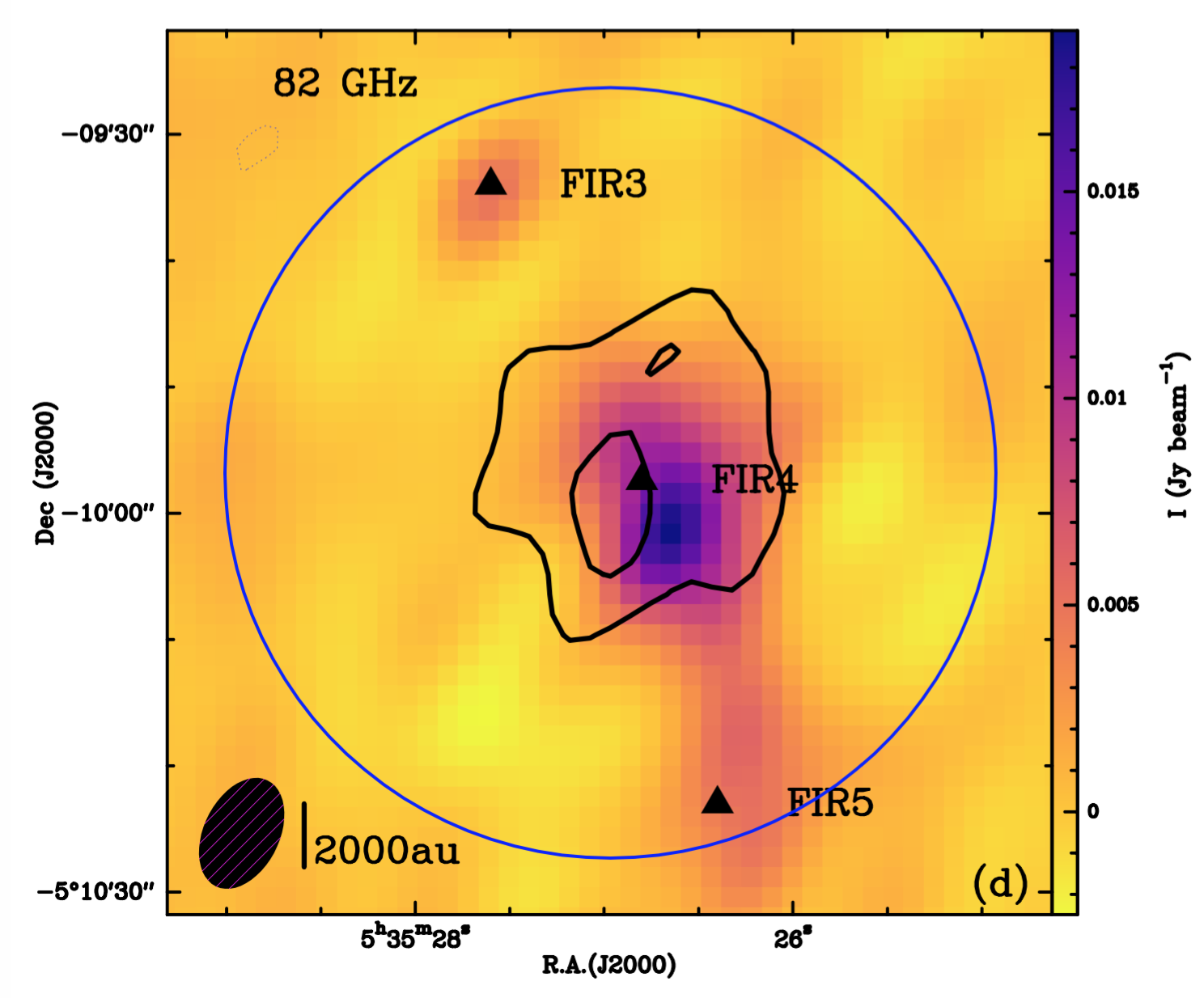}
\caption{ {\it Panel a}: c--C$_3$H$_2$ (2$_{0,2}$--1$_{1,1}$)
  integrated intensity emission map (between 0 and
  22~km~s$^{-1}$). The contour levels are at 3$\sigma$ and 6$\sigma$ (where
  1$\sigma$=2.76$\times$10$^{-3}$~Jy~beam$^{-1}$~km~s$^{-1}$). {\it Panel b}:
  c--C$_3$H$_2$ (3$_{1,2}$--3$_{0,3}$) integrated intensity emission
  map (between 0 and 22~km~s$^{-1}$). The contour levels are at
  3$\sigma$ and 6$\sigma$ (where
  1$\sigma$=2.4$\times$10$^{-3}$~Jy~beam$^{-1}$~km~s$^{-1}$). {\it Panel c}:
  c--C$_3$H$_2$ (2$_{1,2}$--1$_{0,1}$) integrated intensity emission
  map (between 0 and 22~km~s$^{-1}$). The contour levels are at
  3$\sigma$, 4$\sigma$ and 5$\sigma$ (where
  1$\sigma$=2.8$\times$10$^{-3}$~Jy~beam$^{-1}$~km~s$^{-1}$).  {\it Panel d}:
  82~GHz continuum emission (color) overlaid with the integrated
  intensity emission map of the c--C$_3$H$_2$ (3$_{1,2}$--3$_{0,3}$)
  at 82966~MHz (black contours). The contour levels for the continuum and c--C$_3$H$_2$ emission maps are the
  same as in \citet{Fontani:2017} and in {\it Panel b} . Finally, in {\it Panels a, b, c and d}, the
  black triangles indicate the positions of the FIR3, FIR4 and FIR5
  regions \citep[see][]{Chini:1997}; and the blue circle shows the NOEMA field of view.}
\label{fg1}
\end{figure*}

The NOEMA emission maps of the three c--C$_3$H$_2$ lines integrated over the
line profile are shown in Figure~\ref{fg1} (panels a to c). The figure
also displays the continuum emission (panel d), previously reported by \citet{Fontani:2017}, for
reference.

c--C$_3$H$_2$ line emission is detected around FIR4, while FIR 3 and
FIR5 do not show any emission above the 3$\sigma$ level. The emission at 82 GHz
towards FIR4 is rather extended with a hint that it could be
associated with two compact sources north-west and south-east of
FIR4, respectively.
The map at 85 GHz, obtained with a higher spatial resolution, reveals
emission in the same region as the one seen with the 82 GHz lines. Again, the emission
is slightly clumpy (with 1$\sigma$ difference between clumps).  A
forthcoming study, using higher spatial resolution continuum
observations will address the level of core fragmentation in detail
(Neri et al. in preparation).

\begin{figure*}
\centering
\includegraphics[angle=0,width=8.cm]{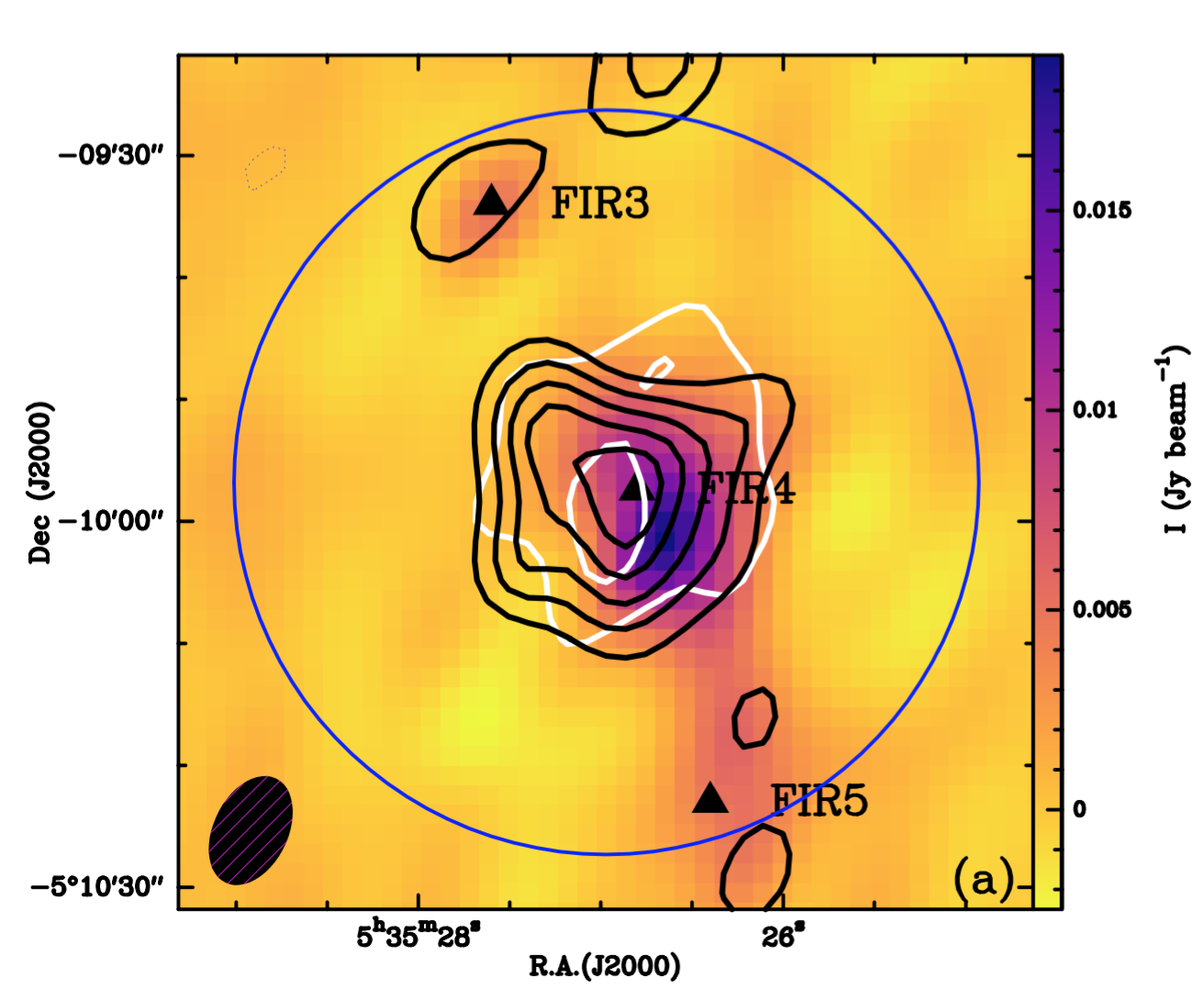} 
\includegraphics[angle=0,width=8.cm]{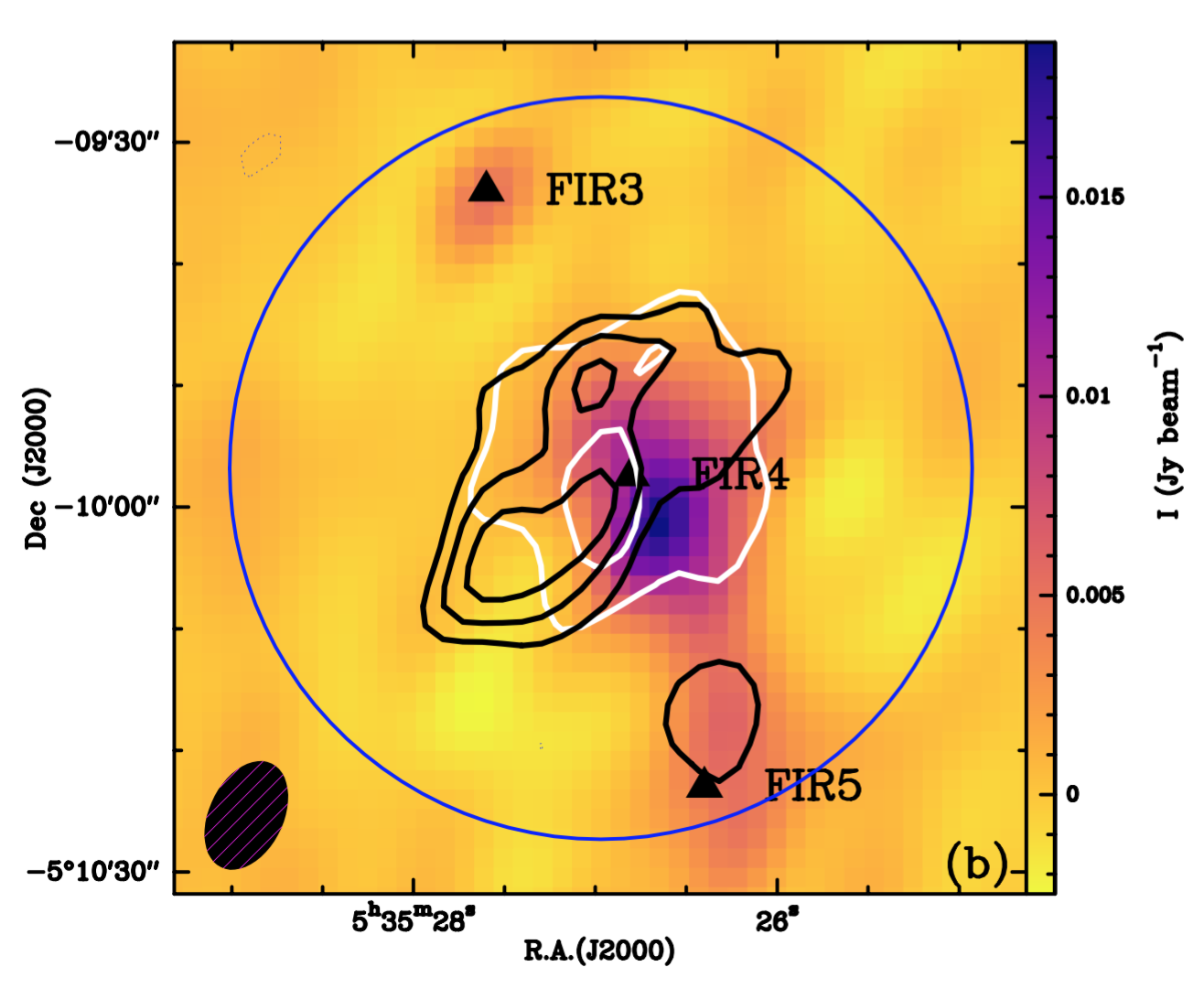}

\caption{{\it Panel a}: 82~GHz continuum emission (color) overlaid with the integrated
  intensity emission map of the c--C$_3$H$_2$ (3$_{1,2}$--3$_{0,3}$)
  at 82.97~GHz (white contours) and that of the HC$_3$N (9--8) at 81.88~GHz  \citep[black contours; from][]{Fontani:2017}.
{\it Panel b} 82~GHz continuum emission (color) overlaid with the integrated
  intensity emission map of the c--C$_3$H$_2$ (3$_{1,2}$--3$_{0,3}$)
  at 82.97~GHz (white contours) and that of the HC$_5$N (31--30) at 82.54~GHz  \citep[black contours; from][]{Fontani:2017}.
Finally, in {\it Panels a and b},  the contour levels for the continuum, HC$_3$N and HC$_5$N maps are the same as in \citet{Fontani:2017}. Contour levels for the c--C$_3$H$_2$ emission map are given in Figure~\ref{fg1}. The blue circle shows the NOEMA field of view.}
\label{fg2}
\end{figure*}

Interestingly enough, the extended c--C$_3$H$_2$ 82 GHz emission is
roughly co-spatial with that of HC$_5$N, \citep[which was observed within the same frequency setting][]{Fontani:2017}, rather than that of HC$_3$N as shown in Figure~\ref{fg2}.

\subsection{c--C$_3$H$_2$ single-dish emission}
\label{extended}

The 30m observations detected 24 c--C$_3$H$_2$ lines, 14 from the
ortho form and 10 from para. Spectra of the c-C$_3$H$_2$ transitions observed with the IRAM 30-m telescope towards OMC--2 FIR\,4 are displayed in Figures \ref{fig:a1a}, \ref{fig:a1b} and \ref{fig:a1c} in Appendix~\ref{appA}. Their properties are reported in Table
\ref{tab:iram30m}. The lines are peaked around the ambient cloud
velocity ($\sim$11 km s$^{-1}$) and are narrow (FWHM$\sim$1.0--1.6
km/s), indicating that they are emitted by the  dense envelope
surrounding FIR4 (see Sec.~\ref{sec:temperature-c-c_3h_2}).
\begin{table}
  \caption{\label{tab:iram30m} c--C$_3$H$_2$ lines detected with the IRAM 30m telescope. }
  \begin{tabular}{rrccc}
    \hline
    \hline
    Freq. & E$_{up}$  & A  & Intensity$^a$ & FWHM$^c$ \\
    (GHz)&	(K)  &  (10$^{-5}$ s$^{-1}$) & K km/s & km/s\\
    \hline
   80.7232  &  28.8   & 1.5     & 0.058  $\pm$   0.006 &     1.6  \\
   {\bf 82.0936$^b$}   &   6.4  & 2.1     & 0.310  $\pm$   0.030  &    1.6    \\
   {\bf 82.9662$^b$}   &  16.0   & 1.1    &  0.290   $\pm$   0.030  &    1.5   \\
   84.7277   &  16.1   & 1.2    &  0.100   $\pm$   0.010   &   1.5  \\
    {\bf 85.3389$^b$}   &  6.4   & 2.6     & 1.200    $\pm$  0.120    &  1.4   \\
   85.6564   &  29.1  & 1.7     & 0.120    $\pm$  0.010    &  1.4  \\
  150.4365  &    9.7   & 5.9     & 0.240    $\pm$  0.050    &  1.2   \\
  150.8207  &   19.3 & 18.0    &  0.590   $\pm$   0.120   &   1.2   \\
  150.8519  &    19.3  & 18 0    &  1.700   $\pm$   0.340   &   1.2   \\
  151.0392  &    54.7  & 6.9     & 0.043    $\pm$  0.009    &  1.1  \\
  151.3439  &    35.4    & 4.4    &  0.110   $\pm$   0.020   &   1.1   \\
  151.3611  &    35.4 & 4.4    &  0.042   $\pm$   0.008   &   1.1   \\
  155.5183  &    16.1   & 12.3    &  0.400   $\pm$  0.080    &   1.1     \\
  204.7889  &    28.8   & 13.7    &  0.130   $\pm$  0.030    &   1.1    \\
  216.2788  &   19.5 & 28.1    &  0.740   $\pm$  0.150   &   1.0    \\
  217.8221  &   38.6  & 59.3    &  1.360   $\pm$  0.270    &   1.2   \\
  217.9400  &   35.4  & 44.3    &  0.700   $\pm$  0.140    &   1.1    \\
  218.1604  &   35.4  & 44.4    &  0.270   $\pm$  0.050    &   1.1     \\
  227.1691  &   29.1   & 34.2    &  0.630   $\pm$  0.120    &  1.0    \\
  244.2221  &   18.2  &   6.5    &  0.230   $\pm$  0.050    &  1.2    \\
  249.0544  &   41.0 & 45.7    &  0.280   $\pm$  0.060    &  1.1    \\
  251.3143  &   50.7   & 93.5    &  0.870   $\pm$  0.170    &  1.2    \\
  254.9876  &   41.1 & 51.7    &  0.120   $\pm$  0.020    &  1.0    \\
  260.4797  &   44.7  & 17.7    &  0.080   $\pm$  0.020    &  1.1     \\
    \hline
    \hline
  \end{tabular}
  {\small
    We used the spectroscopic data parameters from \citet{Bogey:1986},
    \citet{Vrtilek:1987}, \citet{Lovas:1992} and \citet{Spezzano:2012},
    that are available from the Cologne Database for Molecular
    Spectroscopy molecular line catalog
    \citep[CDMS,][]{Muller:2005}. The Einstein coefficients assume an
    ortho-to-para ratio of 3:1. }\\
  {\small NOTES: $^a$The error in the intensity is the quadratic sum of the statistical
    and calibration errors. $^b$The emission from this line was imaged by NOEMA. $^c$The FWHM result from gaussian fit.
    }
\end{table}

\subsection{Missing flux}
\label{missingflux}
To estimate the fraction of the total flux that is filtered out in our
interferometric data, we compared the NOEMA to the IRAM 30m lines. More specifically,
the NOEMA spectra were convolved with a Gaussian beam similar to that
of the 30m beam (i.e. $\sim$30$\arcsec$ at 82--83 GHz). The
convolution was performed at the central position of the IRAM 30m
observations (see Sec. \ref{sec:observations}). Finally, the IRAM-30m
spectra were smoothed to the same spectral resolution
($\sim$7.2~km~s$^{-1}$) as that of the NOEMA WideX spectra. The comparison
of the integrated line fluxes shows that 57$\%$ of the c-C$_3$H$_2$
2$_{0,2}$--1$_{1,1}$ emission, 60$\%$ of that of 3$_{1,2}$--3$_{0,3}$
and 80$\%$ of that of 2$_{1,2}$--1$_{0,1}$ lines is resolved out.
Therefore, the c--C$_3$H$_2$ emission detected by
NOEMA very likely probes the densest part of the envelope surrounding
FIR4 and not the ambient cloud.

%

%
\section{Temperature and c-C$_3$H$_2$ column density}\label{sec:temperature-c-c_3h_2}
%
%
\subsection{Average properties of the FIR4 envelope}\label{sec:outer-envel-cloud}
We first derive the average properties of the FIR4 envelope using the
IRAM 30m data. To this end, we first carried out a simple LTE analysis,
and then a non-LTE one.\\

\subsubsection{LTE analysis}

\begin{figure}
\centering
\plotone{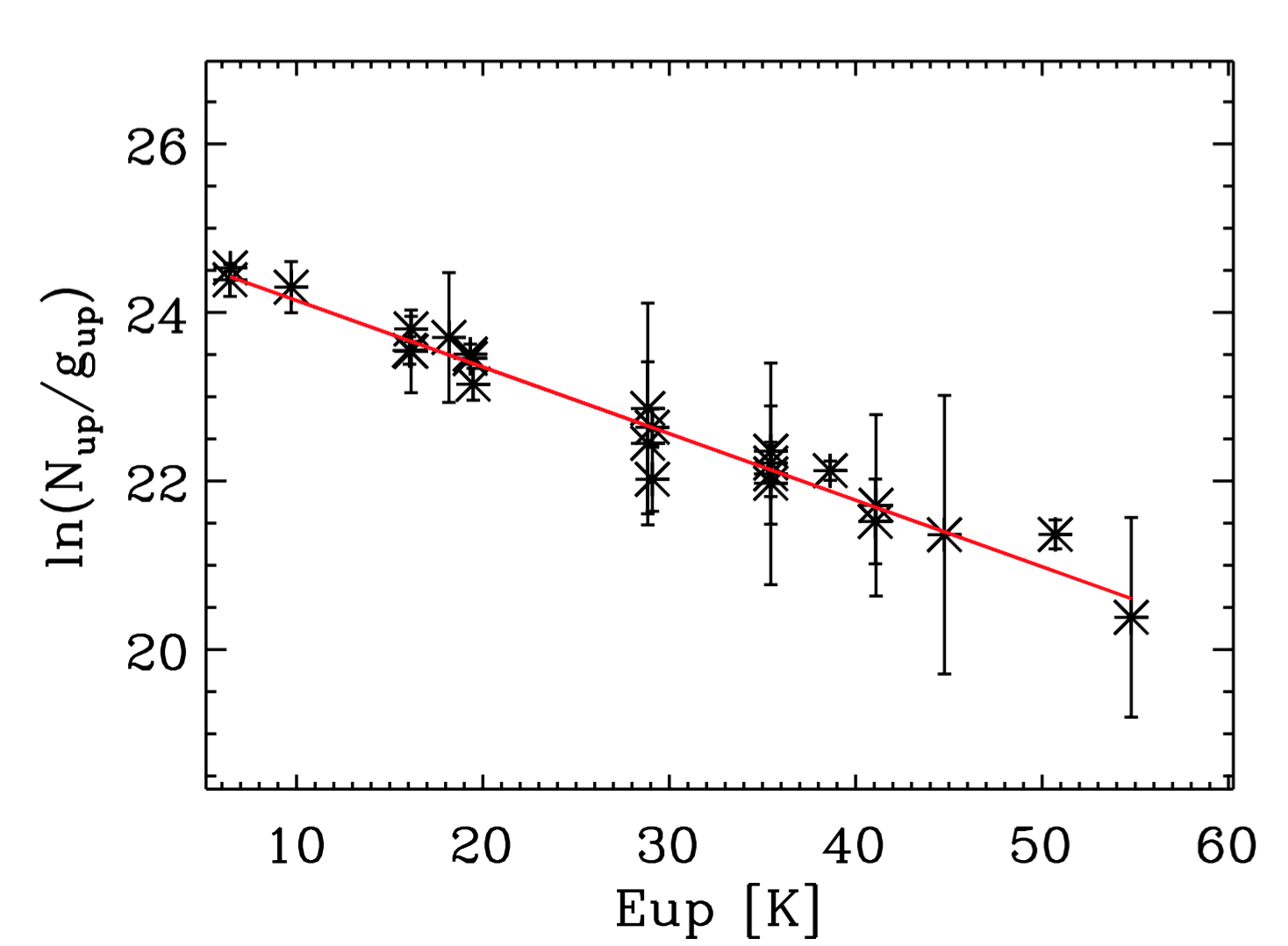}
\caption{Rotational diagram derived from the c-C$_3$H$_2$ lines
  detected with the IRAM 30m telescope (Table \ref{tab:iram30m}). The
  ortho-to-para ratio is taken equal to 3. The error bars correspond to the 1$\sigma$ Gaussian fit uncertainties.}
\label{fg3}
\end{figure}

The rotational diagram, obtained assuming an ortho-to-para ratio equal
to 3 and a beam filling factor equal to 1, is shown in Fig. \ref{fg3}. No systematic difference is seen between ortho and para
lines, which implies that our assumption on the ortho-to-para ratio is
basically correct.
%

The derived rotational temperature is (12.6$\pm$0.5) K and the
c-C$_3$H$_2$ column density is (7$\pm$1)$\times10^{12}$ cm$^{-2}$.
Assuming that the emission arises from a 20$''$ region (see below),
would not change much these results: it would give (10.7$\pm$0.5) K
and (1.5$\pm$0.2)$\times10^{13}$ cm$^{-2}$, respectively, and a
slightly worse (but not significantly better) $\chi^2_{red}$ (0.08
instead of 0.05).\\

\subsubsection{non-LTE analysis}

Given the large number of c-C$_3$H$_2$ lines, we carried out a non-LTE
analysis assuming the Large Velocity Gradient (LVG) approximation. To
this end, we used the LVG code described in \citet{Ceccarelli:2002}
and used the collisional coefficients with He, after scaling for the
different mass of H$_2$, computed by \citet{Chandra:2000} and retrieved
from the BASECOL database\footnote{{\it http://basecol.obspm.fr}: \citet{Dubernet:2013}.}. We assumed a c-C$_3$H$_2$ ortho-to-para
ratio equal to 3, as suggested by the LTE analysis.

We ran a large grid of models with H$_2$ density between $3\times10^4$
and $5\times 10^6$ cm$^{-3}$, temperature between 10 and 50 K and
c-C$_3$H$_2$ column density between $3\times10^{12}$ and
$2\times10^{13}$ cm$^{-2}$. We adopted a FWHM of 1.3 km s$^{-ˆ'1}$ and
let the filling factor be a free parameter. We then compared the
LVG predictions with the observations and used the standard minimum
reduced $\chi^2$ criterium to constrain the four parameters: H$_2$
density, temperature, column density and size. In practice, for each
column density we found the minimum $\chi^2$ in the
density-temperature-size parameter space. The solution then is the one
with the c-C$_3$H$_2$ column density giving the smallest $\chi^2$.

The best fit is obtained for an extended source (i.e. $\geq 30''$), c-C$_3$H$_2$ column density equal to (7$\pm$1)$\times 10^{12}$ cm$^{-2}$ (in excellent agreement with the LTE estimate), temperature equal to 40 K and density equal to $3\times 10^5$ cm$^{-3}$. 
At 2$\sigma$ level, the solution becomes degenerate in the density-temperature parameter space. A family of solutions are possible, with the two extremes at 15 K and $5\times10^6$ cm$^{-3}$ on one side, and 50 K and $2\times 10^5$ cm$^{-3}$ on the other side. 
Please note that, indeed, the coldest and densest solution provides a temperature close to the rotational temperature (13 K). This means that the apparent LTE distribution of the transition levels derived by the IRAM 30m line intensities is also obtained with non-LTE conditions and the densities and temperatures mentioned above, included the best fit solution. Finally, in all cases, the lines are optically thin.
In the following, we will use the best fit solution and we will associate the errors as follows: (40$\pm$10) K and ($3\pm1$) $\times 10^5$ cm$^{-3}$.

Assuming a H$_2$ column density measured from the continuum by
\citet{Fontani:2017}, $\sim 10^{24}$ cm$^{-2}$, we obtain an average
c-C$_3$H$_2$ abundance of (7$\pm$1)$\times10^{-12}$, assuming that
c-C$_3$H$_2$ is present across the entire OMC--2 FIR\,4 region (Section
\ref{sec:chemical-modeling}).

\subsection{The structure of the FIR4 envelope}\label{sec:fir4-env}

The maps obtained with the NOEMA interferometer allow us to estimate
the gas temperature and the c-C$_3$H$_2$ abundance across the region
probed by the NOEMA observations.

\begin{figure}
\centering
\includegraphics[angle=0,width=\hsize]{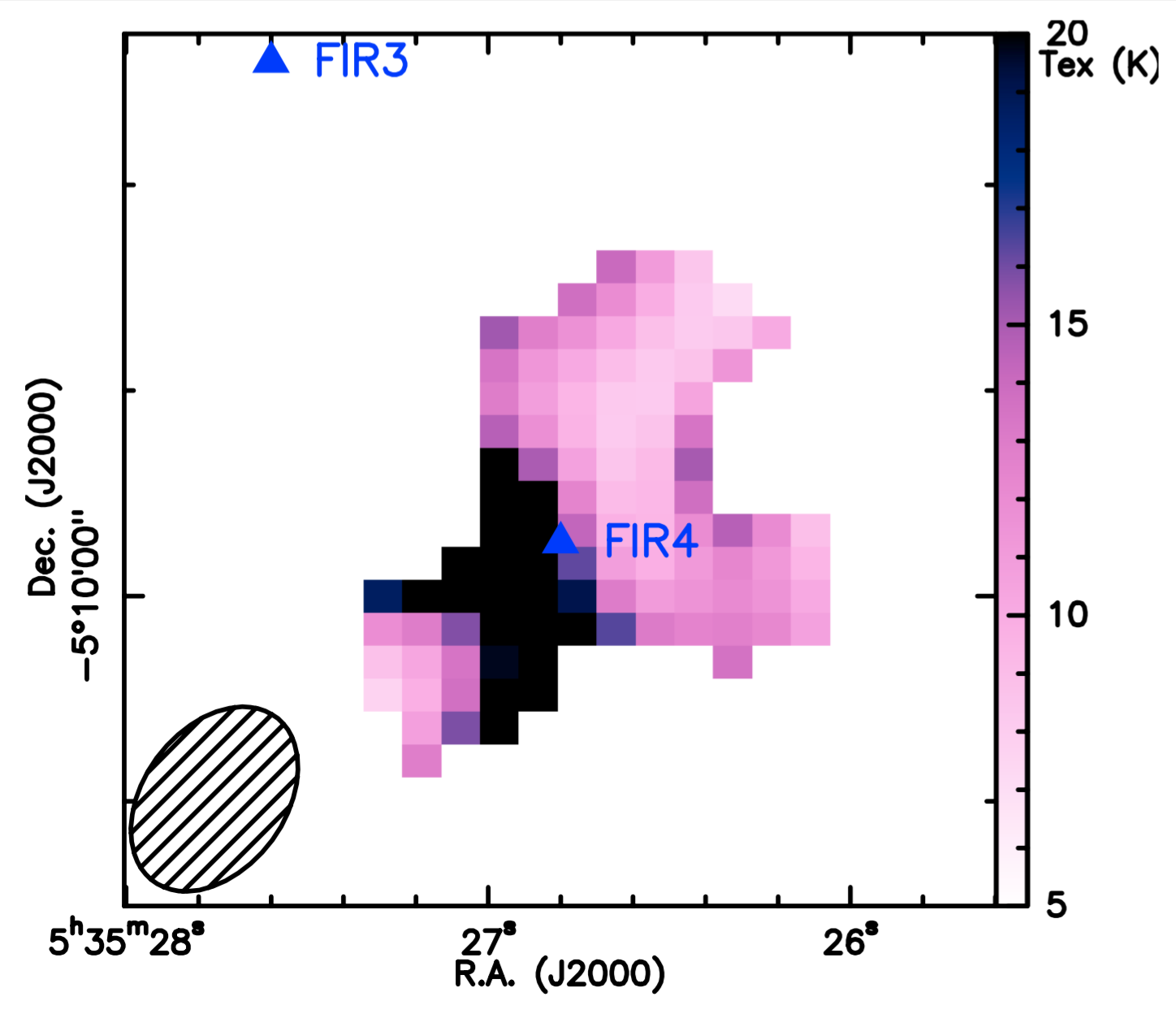} 
\includegraphics[angle=0,width=\hsize]{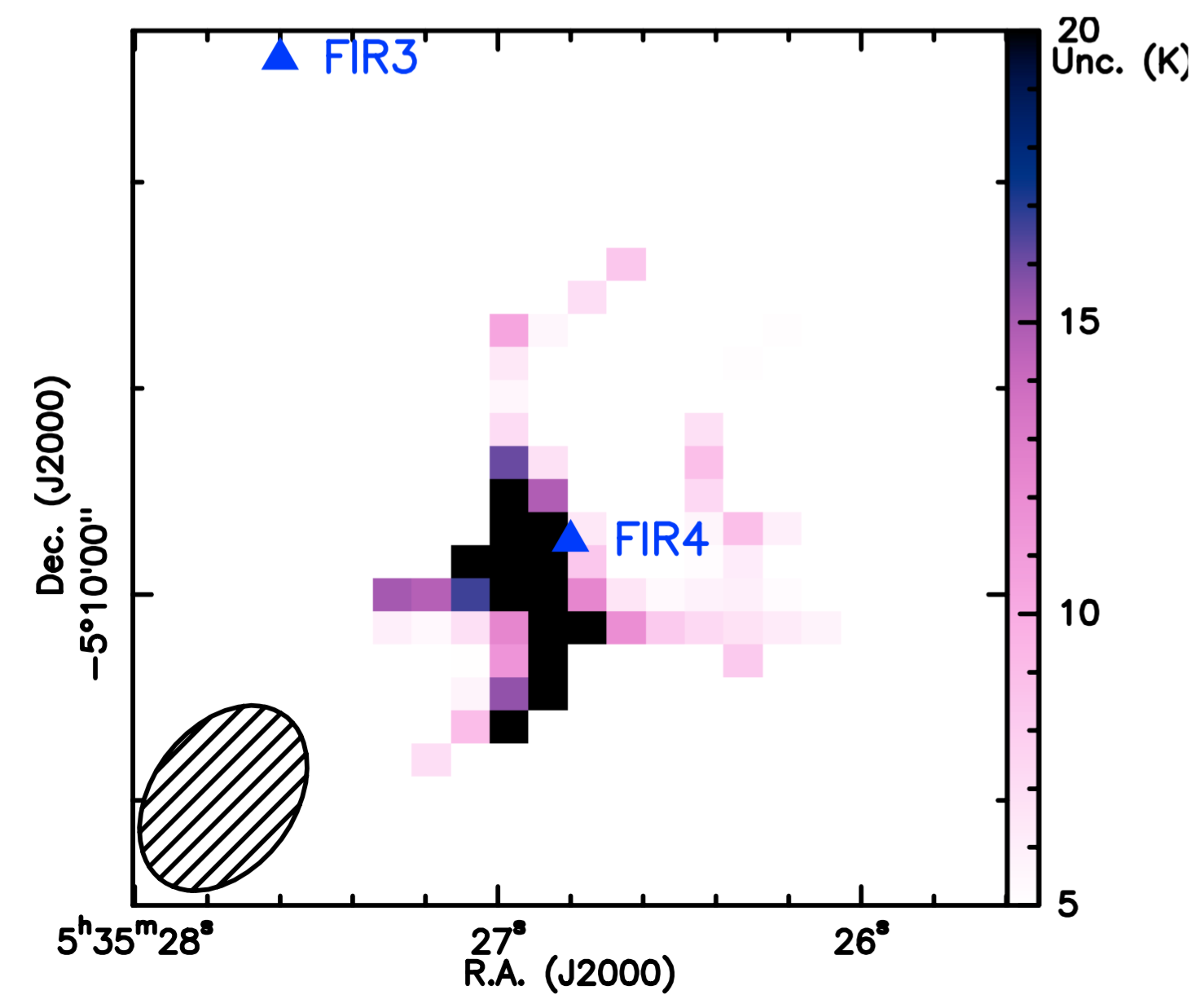} 
\caption{ c--C$_3$H$_2$ excitation temperature map towards OMC2--FIR4
  (top panel) along with the uncertainty map (bottom panel). }
\label{fg4}
\end{figure}

To measure the excitation temperature, we use the two c-C$_3$H$_2$ lines at
82 GHz, which have the same spatial resolution and, most importantly,
the same amount of filtered out extended emission\footnote{Please note
  that the 85 GHz line has a filtering twice larger than that of the
  lines at 82 GHz (Sec.  \ref{missingflux}).}.  To this end, we
assumed that (1) the c--C$_3$H$_2$ ortho-to-para ratio is equal to 3,
(2) the lines are optically thin, and (3) the levels are LTE populated. With this
assumptions, we derived the excitation temperature map shown in Figure
\ref{fg4}, along with the associated uncertainty. The excitation
temperature is comprised between 8$\pm$3 and 16$\pm$7~K (i.e. minimum and maximum values, see Fig.~\ref{fg4}). When
considering the uncertainty on the values, there are no signs of
excitation temperature gradients across the region we are probing (about 3-4 beams). On the contrary, the
excitation temperature seems to be rather constant and not much different
from that probed by the 30m data analysis ($\sim$12 K) (see previous
section).

%

\section{Chemical modeling}\label{sec:chemical-modeling}
In the previous Section, we showed that the gas emitting the
c--C$_3$H$_2$ lines has a temperature of $\sim 40$ K and H$_2$ density
$\sim 3\times10^5$ cm$^{-3}$. The c--C$_3$H$_2$ column density is
$\sim 7\times10^{12}$ cm$^{-2}$.
In this section, we use a Photo-Dissociation Region (PDR) model to
understand the structure of the gas probed by the c--C$_3$H$_2$ lines,
notably where they come from, and what constraints they provide.

\begin{figure}
\centering
  \includegraphics[angle=0,width=9cm]{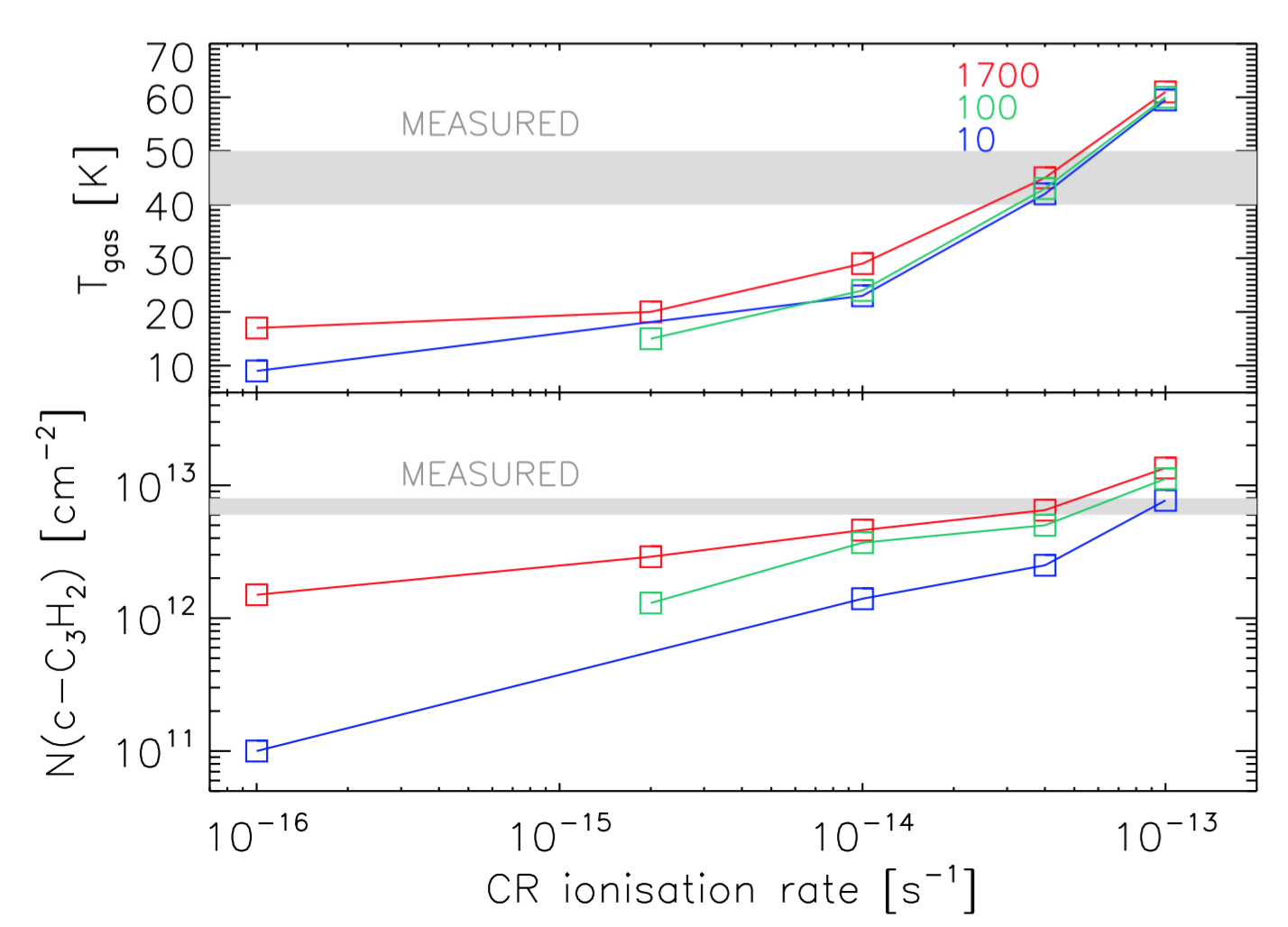}
  \caption{Results of the PDR modeling: predicted c-C$_3$H$_2$ column
    density (lower panel) and gas temperature (upper panel) as a
    function of the CR ionisation rate $\zeta_{CR}$, for a FUV field
    $G_0$ equal to 1700 (red), 100 (green) and 10 (blue), respectively. Symbols
    show the actual runs. }
\label{fig:pdr-model}
\end{figure}

\begin{figure}
  \includegraphics[angle=0,width=8cm]{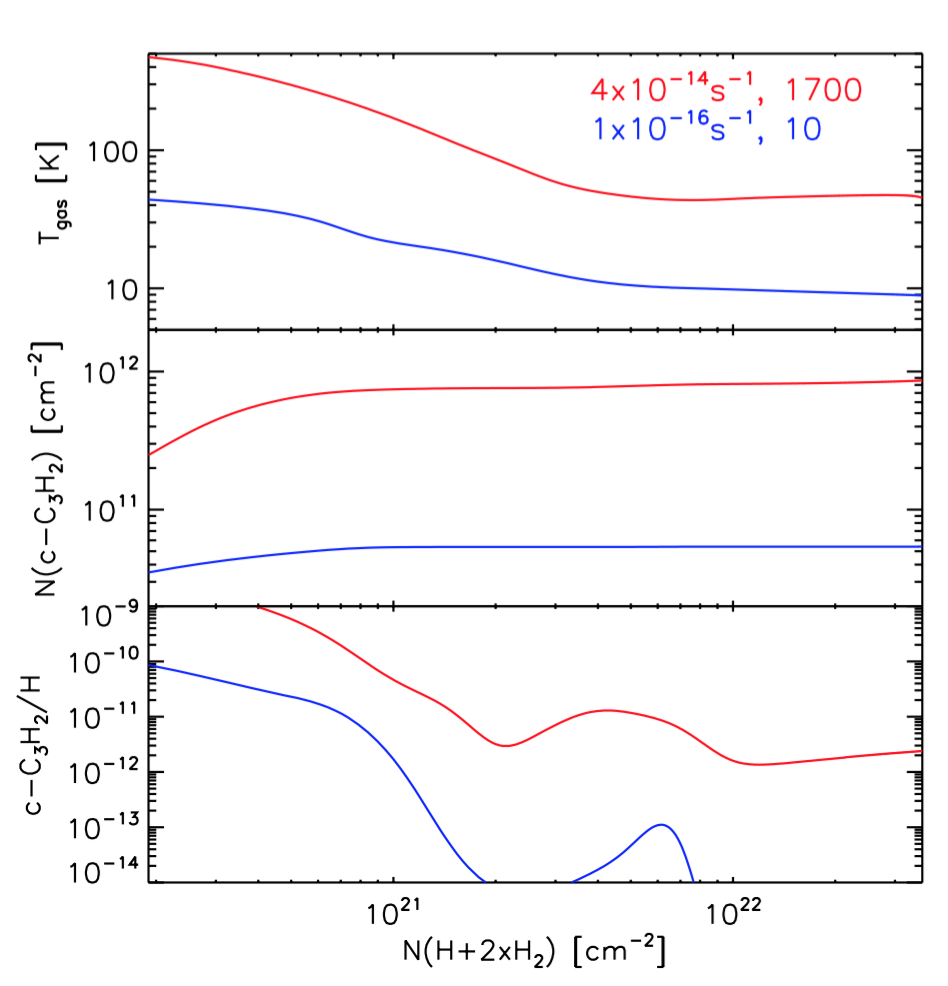}
  \caption{Thermal and chemical structure of two models: predicted gas
    temperature (upper panel), c-C$_3$H$_2$ column density in the PDR
    (middle panel) and abundance (bottom panel) as a function of the H
    nuclei column density. Two models are reported: model 12 ($\zeta_{CR}=4\times 10^{-14}$ s$^{-1}$ and $G_0=1700$; red), and
    model 2 ($\zeta_{CR}=1\times 10^{-16}$ s$^{-1}$ and $G_0=10$; blue).  }
\label{fig:pdr-struct}
\end{figure}

To this end, we used the Meudon PDR code\footnote{The code is publicly
  available at http://pdr.obspm.fr.} \citep[version 1.5.2, see][]{Le-Petit:2006,Bron:2016}. The code computes the steady-state thermal and chemical
structure of a cloud irradiated by a FUV radiation field $G_o$ and
permeated by CR that ionise the gas at a rate $\zeta_{CR}$. Relevant
to the chemistry, the code computes the gas-phase abundances of the
most abundant species, including c--C$_3$H$_2$.
In our simulations, we adopted the density derived by the
c--C$_3$H$_2$ non-LTE analysis ($n_H=6\times10^5$ cm$^{-3}$) and the elemental abundances listed in Table \ref{tab:ele_abu}.  
Note that we
limited the PDR simulations to a H-nuclei column density N$_H$ of
$\sim 4\times10^{22}$ cm$^{-2}$ (corresponding to A$_v$=20 mag; see
Figs. \ref{fig:pdr-model} and \ref{fig:pdr-struct}). However, to compare the final predicted
c--C$_3$H$_2$ column density, N$_{Tot}$, with the observed one
($7\times10^{12}$ cm$^{-2}$), we have to consider the whole cloud,
which has a total H-nuclei column density N$_H^{FIR4}$ of
$2\times10^{24}$ cm$^{-2}$ \citep{Fontani:2017}. Therefore, we
multiplied the c--C$_3$H$_2$ abundance predicted by the model in the
cloud interior X$_{interior}$ (computed by the code at
N$_H=3\times10^{22}$ cm$^{-2}$)\footnote{Please note that we used the
  c--C$_3$H$_2$ abundance at N$_H=3\times10^{22}$ cm$^{-2}$ to avoid
  the region where opacities and, consequently, temperatures decrease
  because of the artificial boundary of the cloud.} by N$_H^{FIR4}$,
and added it to the c--C$_3$H$_2$ column density in the PDR region
N$_{PDR}$ (computed by the code for N$_H\leq 1\times10^{22}$
cm$^{-2}$), as follows:
\begin{equation}
{\rm N}_{Tot} = {\rm X}_{interior}\times {\rm N}_H^{FIR4} + 2\times {\rm N}_{PDR}
\end{equation}

\begin{table}
  \begin{center}
  \begin{tabular}{cc|cc}
    \hline
    \hline
    Element & Abundance & Element & Abundance \\
    \hline
    O & $3.2\times10^{-4}$ & C & $1.3\times10^{-4}$\\
    N & $7.5\times10^{-5}$ & S & $1.9\times10^{-5}$\\
    Si & $8.2\times10^{-7}$ & Fe & $1.5\times10^{-8}$\\
    \hline
    \hline
  \end{tabular}
  \caption{Assumed elementary abundances, with respect to H nulei, in
    the PDR modeling. }
      \end{center} 
  \label{tab:ele_abu}
\end{table}

To initialize our grid of models, we used as input parameter a temperature of  50~K, and assume an edge-on region that is irradiated from one side only. Then, we run several models with values of $G_0$ ($G_0=$ is the FUV radiation field
in Habing units\footnote{$G_0 = 1$ corresponds to a FUV energy density of $5.3\times10^{-14}$
erg cm$^{-3}$. The interstellar standard radiation field is $G_0 = 1.7$.}) 
ranging from 1 to 1700, and $\zeta_{CR}$ from
$1\times 10^{-16}$ s$^{-1}$ to $4\times 10^{-14}$ s$^{-1}$. These
extreme values are quoted in the literature for the OMC--2 region (see
Introduction and Discussion). Note that, even though we did not run a
full grid of models, we fine-tuned the parameter ranges close to the
best fit solution.

\begin{table*}
\centering
 \begin{tabular}{ccc|cccc}
   \hline
   \hline
  Model & $\zeta_{CR}$ & $G_0$ & N$_{PDR}$ & X$_{interior}$ & N$_{Tot}$  & T$_{gas}$\\
           & ($10^{-16}$ s$^{-1}$) & & ($10^{12}$ cm$^{-2}$) & & ($10^{12}$ cm$^{-2}$) & (K)\\
   \hline
   1  & 1     & 1         & 0.01 & $9.8\times 10^{-16}$ & 0.02 & 9\\
   2  & 1     & 10       & 0.05 & $1.5\times 10^{-15}$ & 0.1 & 9\\
   3  & 1     & 1700   & 0.75 & $2.5\times 10^{-14}$ & 1.5 & 17\\
   4  & 20   & 100     & 0.11  & $5.3\times 10^{-13}$ & 1.3 & 14\\
   5  & 20   & 1700   & 0.77  & $7.1\times 10^{-13}$ & 2.9 & 20\\
   6  & 100 & 10       & 0.12 & $5.7\times 10^{-13}$ & 1.4 & 23\\
   7  & 100 & 100     & 0.18 & $1.6\times 10^{-12}$ & 3.7 & 25\\
   8  & 100 & 1700   & 0.79 & $1.5\times 10^{-12}$ & 4.6 & 29\\
   9  & 400 & 1        & 0.06 & $7.4\times 10^{-13}$ & 1.6 & 41\\
  10 & 400 & 10      & 0.19 & $1.1\times 10^{-12}$ & 2.5 & 42\\
  11 & 400 & 100    & 0.26 & $2.2\times 10^{-12}$ & 5.0 & 43\\
  12 & 400 & 1700  & 0.86  & $2.4\times 10^{-12}$ & 6.5 & 45\\
   \hline
   \hline
 \end{tabular}
 \caption{List of the $\zeta_{CR}$ and $G_0$ values adopted for the different PDR
   models and the results of the simulations: the c-C$_3$H$_2$ column
   density in the PDR region, N$_{PDR}$; the
   c-C$_3$H$_2$ abundance (with respect to H nuclei) in the inetrior
   (see text), X$_{interior}$; the total c-C$_3$H$_2$ column density,
   N$_{Tot}$, which takes into account the 
   whole cloud (see text); the gas temperature (in the
   interior). }
 \label{tab:pdr-model}
\end{table*}
%


The run models and the associated results are listed in Table
\ref{tab:pdr-model} and shown in Figure \ref{fig:pdr-model}.  Both
the predicted c-C$_3$H$_2$ column density and gas temperature are
strong functions of $\zeta_{CR}$: the larger $\zeta_{CR}$ the larger
the column density and the temperature. On the contrary, the
value of $G_0$ has a small influence on the predicted values, in particular for the temperature.

The comparison of the PDR modeling results (Table \ref{tab:pdr-model}
and Figure \ref{fig:pdr-model}) with the measured c-C$_3$H$_2$ column
density and gas temperature very clearly indicates that the gas is
permeated by a large flux of CR and is irradiated by an intense FUV
field. Specifically, model 12 ($\zeta_{CR}=4\times 10^{-14}$ s$^{-1}$
and $G_0=1700$) reproduces fairly well the measured c-C$_3$H$_2$
column density ($\sim 7\times10^{12}$ cm$^{-2}$) and gas temperature
($\sim 40$ K).

%
Figure \ref{fig:pdr-struct} shows the gas temperature, the
c-C$_3$H$_2$ abundance and column density as a function of
N(H+2$\times$H$_2$) for the model 12, which best reproduces the
observations, and model 2, for comparison.

In general, the c-C$_3$H$_2$ abundance has a first peak in the PDR
region, in a tiny layer at N(H+2$\times$H$_2$) $\leq \times 10^{21}$
cm$^{-2}$. This first peak depends on the FUV radiation field and
increases with increasing $G_0$. In the interior of the cloud, at
N(H+2$\times$H$_2$) $\geq\times 10^{22}$ cm$^{-2}$, namely in the
region that contributes most to the total c-C$_3$H$_2$ column density,
the c-C$_3$H$_2$ abundance is governed by the CR ionisation rate and
increases with $\zeta_{CR}$. As expected, the gas temperature at the
PDR border is governed by the FUV field, whereas it is governed by CR
ionisation rate in the interior.

%
\section{Discussion}
\label{discussion}
%
\subsection{OMC--2 FIR\,4: a highly irradiated region}\label{sec:omc-2-fir4}

In the previous section, we showed that in order to reproduce the
temperature of the gas probed by the c-C$_3$H$_2$ lines and its
abundance the gas has to be irradiated by a strong flux of CR-like
particles. 
Amazingly, the best agreement between observations and model predictions is given by a CR ionisation rate,
$\zeta_{CR}=4\times 10^{-14}$ s$^{-1}$, that is the same as the one derived
by the following observations: (1) the HCO$^+$ and N$_2$H$^+$ high
J lines observed by Herschel HIFI CHESS project \citep{Ceccarelli:2014}, and (2) the HC$_3$N and HC$_5$N lines observed by NOEMA
SOLIS project \citep{Fontani:2017}. In addition, the c-C$_3$H$_2$ emitting region roughly coincides with the largest $\zeta_{CR}$ region which is probed by the HC$_5$N emission area.
We emphasise that, in addition to
being different data sets and different species, the three estimates of
$\zeta_{CR}$ have been obtained also with three different astrochemical
codes: ASTROCHEM\footnote{http://smaret.github.com/astrochem/}, Nahoon \citep{Wakelam:2012} and Meudon PDR \citep[version 1.5.2,][]{Le-Petit:2006,Bron:2016} codes.

\begin{figure*}
\centering
  \includegraphics[angle=0,width=16cm]{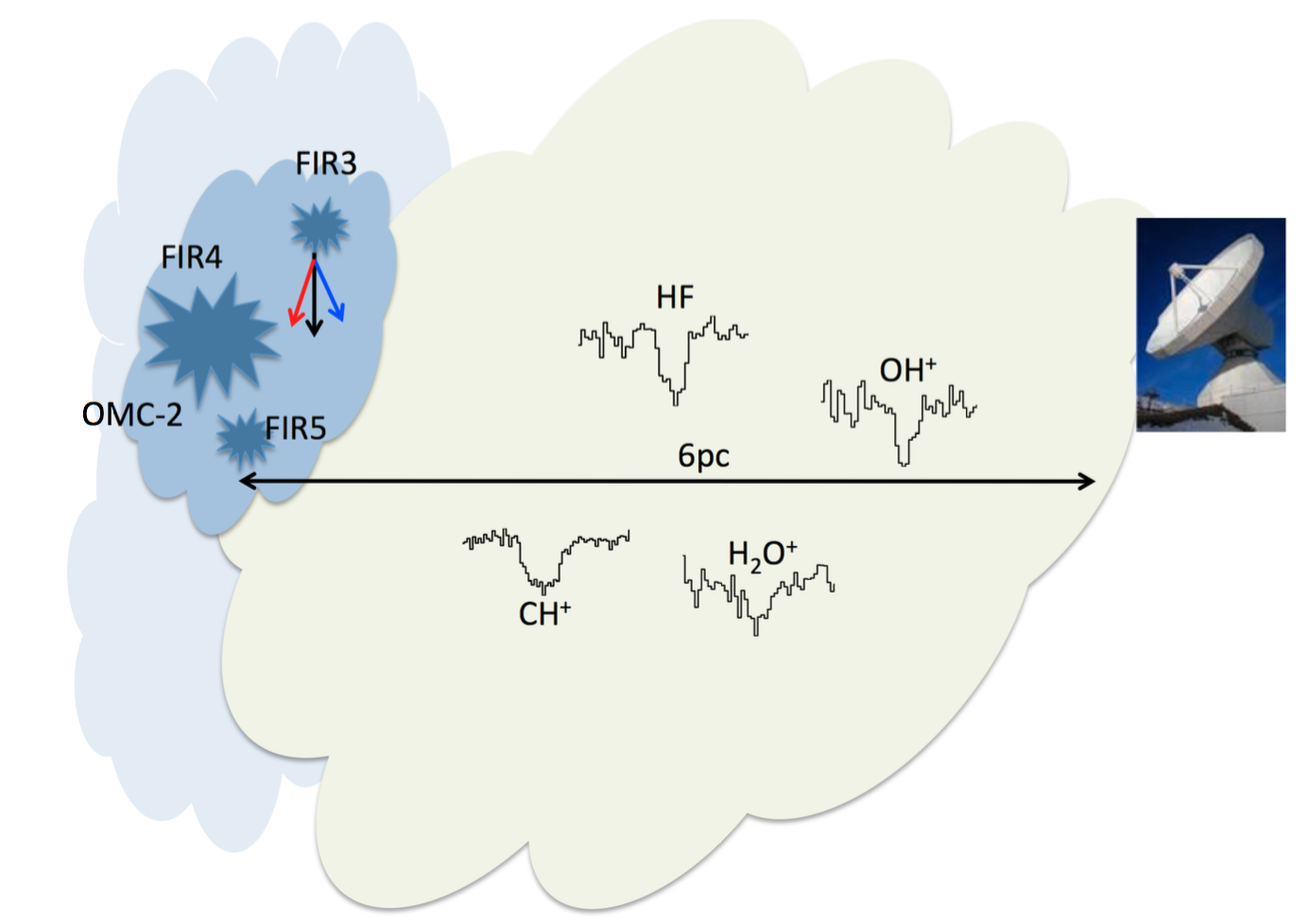}
  \caption{Cartoon of OMC--2.  A tenuous cloud between
    OMC--2 and the Sun is probed by the absorption of HF, OH$^+$,
    CH$^+$ and H$_2$O$^+$ observed by the Herschel HIFI instrument
    \citep{Lopez-Sepulcre:2013a}. The OMC--2 complex is associated with three FIR
    sources: FIR3, FIR4 and FIR5, respectively. An outflow is emitted
    by FIR3 which, based on the present observations, is unlikely
    impacting the FIR4 envelope. }
\label{fig:region}
\end{figure*}

The emerging picture is shown in the cartoon of Figure
\ref{fig:region}. Previous Herschel HIFI CHESS observations showed
that between OMC--2 and us there is a tenuous ($100$ cm$^{-3}$) cloud
extending about 6 pc along the line of sight, and illuminated by a FUV
field about 1000 times brighter than the interstellar standard radiation field \citep{Lopez-Sepulcre:2013a}. 
OMC--2 itself hosts three FIR sources
(FIR 3, 4 and 5), which are likely very different in nature and
evolutionary status, even though not much is known especially about
FIR3 and FIR5, except that a large-scale ($\sim 30''$) outflow
emanates from FIR3 \citep{Takahashi:2008,Shimajiri:2008,Shimajiri:2015}. FIR4 is actually a dense clump \citep[$\sim 10^5-10^6$ cm$^{-3}$:][]{Crimier:2009,Ceccarelli:2014}
 with an embedded cluster of
young sources, whose number is still unclear \citep{Shimajiri:2008,Lopez-Sepulcre:2013,Kainulainen:2017}. What is clear now is that FIR4 is permeated by a flux of CR-like ionising particles about 1000 times larger than the CR flux of the Galaxy. The
source(s) of these particles is(are) likely situated in the East part
of FIR4 \citep{Fontani:2017}, but still remains unidentified. Incidentally, it is important to note that the high CR ionization rate could result in a temperature gradient in the vicinity of the CR emitting source(s). Nonetheless, the c-C$_3$H$_2$ excitation temperature mainly covers the west region (see Figure~\ref{fg4}), where the irradiation is likely lower, based on the HC$_5$N/HC$_3$N abundance ratio by \citet{Fontani:2017}.
 The new IRAM 30m and
SOLIS observations presented in this work confirm this geometry and indicate that abundant hydrocarbons (c-C$_3$H$_2$)
are present not only at the skin of the FIR4 clump but also in the
interior, because of the strong CR-like irradiation.

%

\subsection{No evidence of the FIR3 outflow impact on FIR4}\label{sec:no-evidence-fir3}
It has been suggested that the chemical composition of OMC--2 FIR4 is
affected by the interaction of the northeast--southwest outflow driven
by the nearby source FIR3
\citep[see][]{Shimajiri:2008,Shimajiri:2015}, which is located at
about 23$\arcsec$ northeast from FIR4 \citep[i.e. $\sim$9660~AU at a
distance of 420~pc,][]{Menten:2007,Hirota:2007}. Specifically,
\citet{Shimajiri:2008,Shimajiri:2015} have suggested that gas
associated with OMC--2 FIR4 might be impacted by this NE-SW
outflow. If this is the case, the gas associated with the envelope
of the OMC--2 FIR4 region should show some physically induced
effect. In this instance, the c-C$_3$H$_2$ molecular emission map
would likely present a temperature gradient within the region. This is
not the case in our observations (see Figure \ref{fg4}) which,
contrary to \citet{Shimajiri:2015}, probe the envelope of OMC--2 FIR4
and not the ambient gas thanks to the interferometer spatial
filtering.
These findings lead one to ask whether the apparent spatial coincidence of the
southern outflow lobe driven by FIR3 and the northern edge of FIR4 is
simply a projection effect. More sensitive, higher angular resolution observations may help us confirm our current conclusion.

%
\section{Conclusions}
\label{conclusions}

We have imaged, for the first time, the distribution of cyclopropenylidene, c--C$_3$H$_2$, towards OMC--2 FIR\,4 with an angular resolution of 9.5\arcsec$\times$6.1\arcsec at 82 GHz and 4.7\arcsec$\times$2.2\arcsec at 85 GHz, using NOEMA. The observations were performed as part of the SOLIS program. In addition, we have performed a study of the physical properties of this source through the use of IRAM-30m observations. 

Our main results and conclusions are the following:
\begin{enumerate}
\item From a non-LTE analysis of the IRAM-30m data, we find that c--C$_3$H$_2$ gas emits at the average temperature of about 40~K with a  $\chi$(c--C$_3$H$_2$) abundance of ($7\pm1$)$\times 10^{-12}$.
\item Our NOEMA observations show that there is no sign of excitation temperature gradients within the observed region (which corresponds to $\sim$3-4 beams), with a T$_{ex}$(c--C$_3$H$_2$) in the range 8$\pm$3 -- 16$\pm$7 K. Our findings suggest that the OMC--2 FIR\,4 envelope is not in direct physical interaction with the outflow originating from OMC--2 FIR\,3.
\item In addition, the c--C$_3$H$_2$ gas probed by NOEMA arises from the same region as that of HC$_5$N which is a probe of high CR-particles ionization \citep{Fontani:2017}.
\item Finally, a notable result, derived from chemical modelling with the Meudon PDR code is that OMC--2 FIR\,4 appears to be a strongly irradiated region: FUV field dominates the outer shells (with a radiation field scaling factor $G_0$ of about 1700) while the interior of the envelope is governed by CR ionization (with a CR ionization rate $\zeta_{CR}$ = $4\times10^{-14}$ s$^{-1}$, namely more than a thousand times the canonical value). 
\end{enumerate}
These results are consistent with previous studies claiming that OMC--2 FIR\,4 bathes in an intense radiation field of energetic particles ($\geq10$ MeV).

%
%
%
\section{Acknowledgements}
We warmly thank Franck Le Petit for his assistance in the use of the PDR code.
CF acknowledge the financial support for this work provided by the
French space agency CNES along with the support from the Italian Ministry of Education, Universities and Research, project SIR (RBSI14ZRHR).
We acknowledge the funding from the
European Research Council (ERC), projects PALs (contract 320620) and
DOC (contract 741002).
This work was supported by the French program "Physique et Chimie du
Milieu Interstellaire" (PCMI) funded by the Conseil National de la
Recherche Scientifique (CNRS) and Centre National d'€™Etudes Spatiales
(CNES) 
and by the Italian Ministero dell'Istruzione, Universit\'a e
Ricerca through the grant Progetti Premiali 2012 - iALMA (CUP
C52I13000140001).

%
%
\appendix

\section{c-C$_3$H$_2$ towards OMC--2 FIR\,4 as observed with IRAM-30m telescope}\label{appA}

\begin{figure}
\centering
  \includegraphics[angle=0,width=16cm]{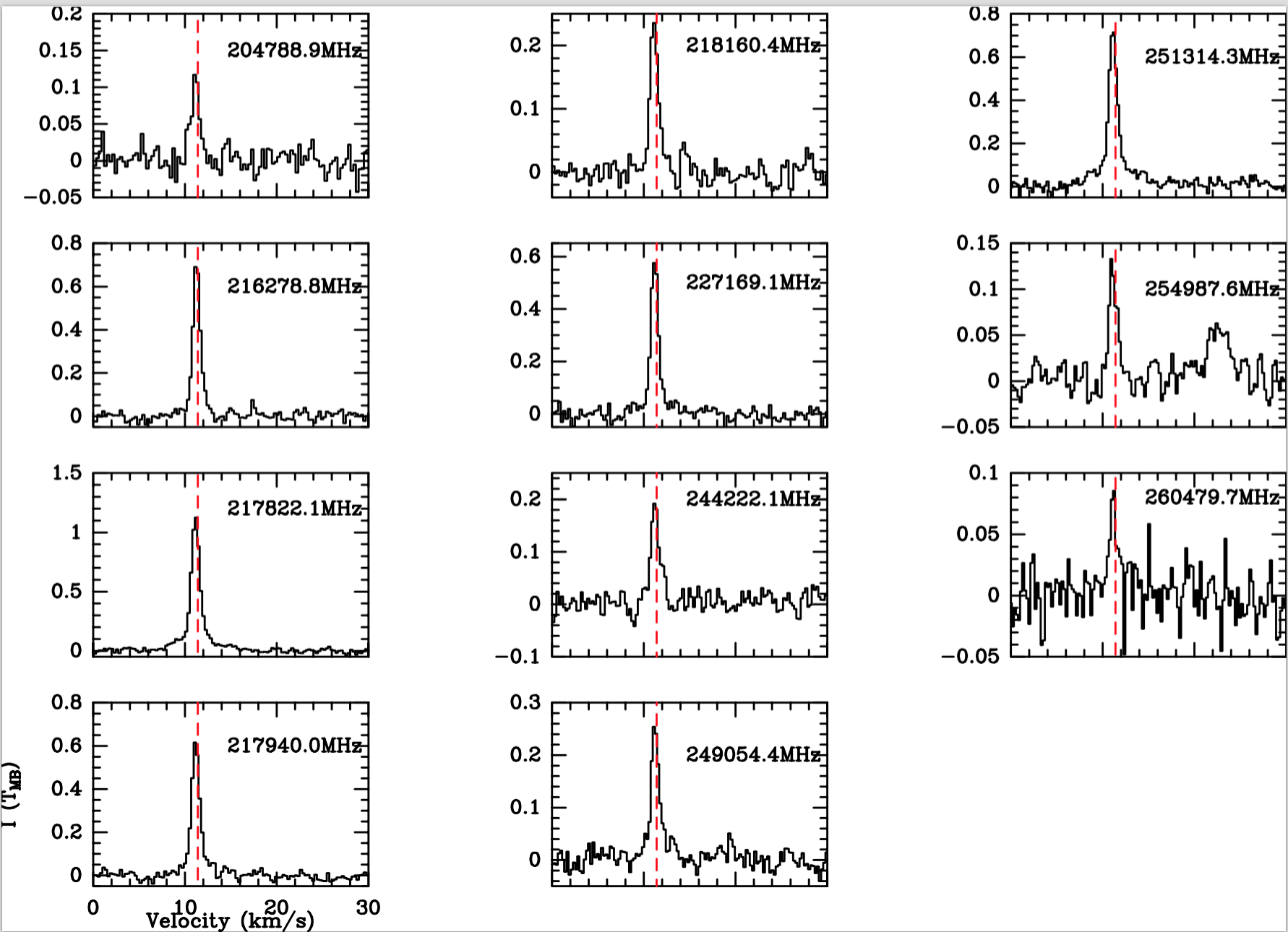}
  \caption{c-C$_3$H$_2$ spectra observed towards OMC--2 FIR\,4 at 1~mm with the IRAM 30-m telescope. Dashed red lines indicate a  $V_{LSR}=11.4$~km~s$^{-1}$. The frequency the observed transition is indicated on each plot.}
\label{fig:a1a}
\end{figure}

\begin{figure}
\centering
  \includegraphics[angle=0,width=16cm]{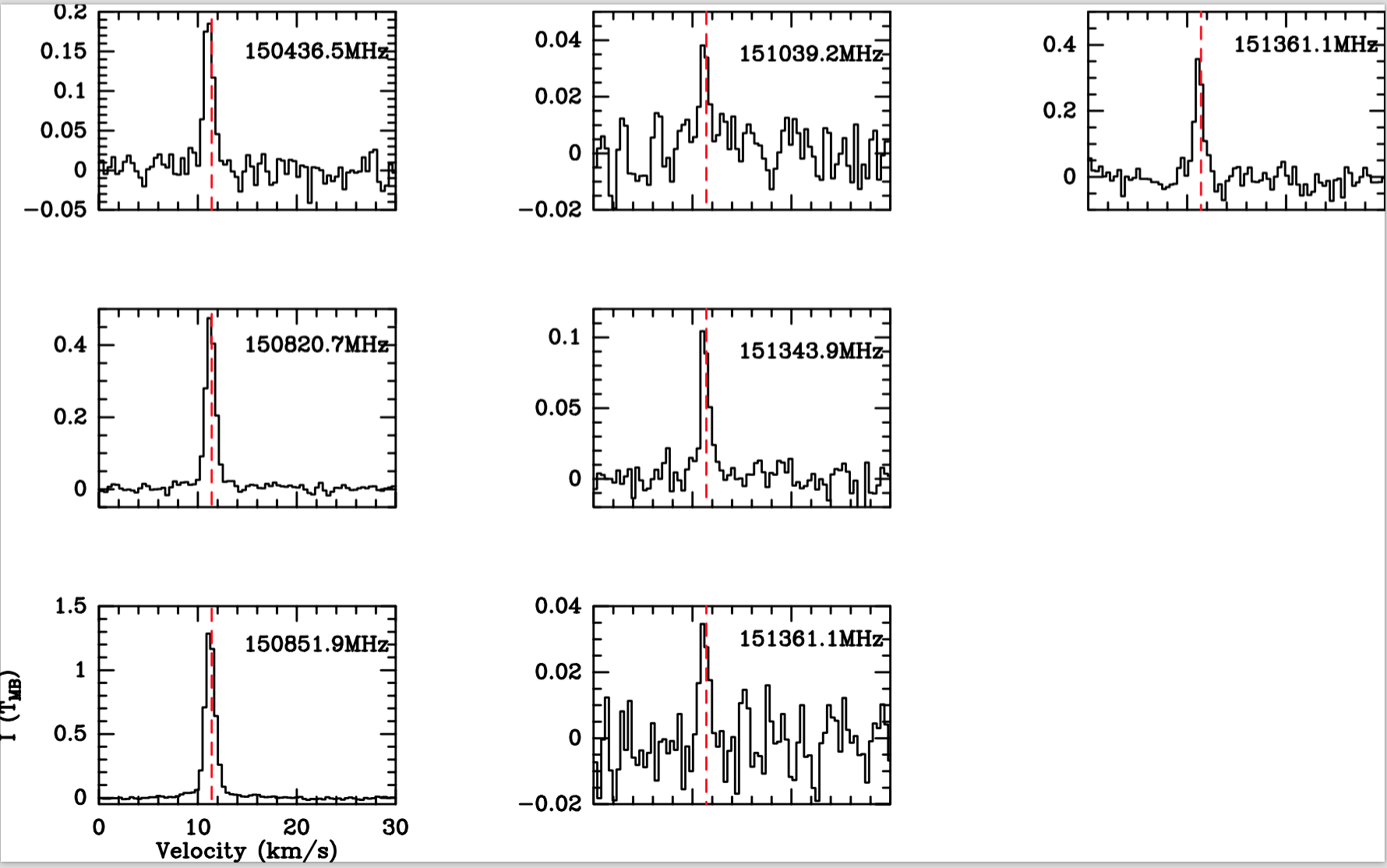}
  \caption{c-C$_3$H$_2$ spectra observed towards OMC--2 FIR\,4 at 2~mm with the IRAM 30-m telescope. Dashed red lines indicate a  $V_{LSR}=11.4$~km~s$^{-1}$. The frequency the observed transition is indicated on each plot.}
\label{fig:a1b}
\end{figure}

\begin{figure}
\centering
  \includegraphics[angle=0,width=16cm]{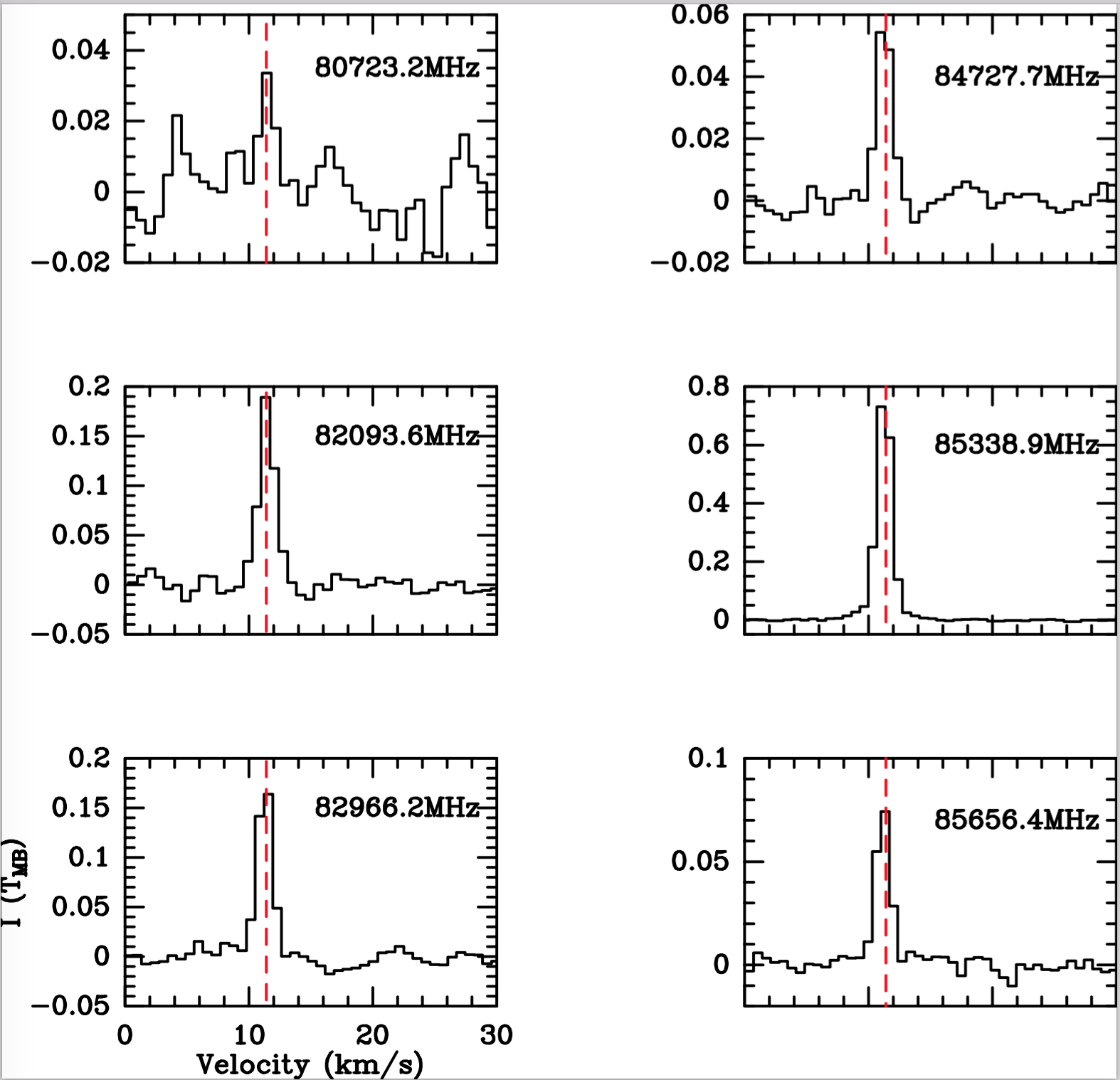}
  \caption{c-C$_3$H$_2$ spectra observed towards OMC--2 FIR\,4 at 3~mm with the IRAM 30-m telescope. Dashed red lines indicate a  $V_{LSR}=11.4$~km~s$^{-1}$. The frequency the observed transition is indicated on each plot.} 
\label{fig:a1c}
\end{figure}

Figures \ref{fig:a1a} to  \ref{fig:a1c} display the spectra of the c-C$_3$H$_2$ transitions observed with the IRAM 30-m telescope towards OMC--2 FIR\,4 at 1, 2 and 3~mm, respectively.

%
%
%

\bibliographystyle{aasjournal}

\end{document}